%% file: main.tex
\documentclass[10pt,conference]{IEEEtran}
\IEEEoverridecommandlockouts

\usepackage{cite}
\usepackage{amsmath,amssymb,amsfonts}
\usepackage{algorithmic}
\usepackage{graphicx}
\usepackage{textcomp}
\usepackage{xcolor}
\usepackage{subcaption}

\def\BibTeX{{\rm B\kern-.05em{\sc i\kern-.025em b}\kern-.08em
T\kern-.1667em\lower.7ex\hbox{E}\kern-.125emX}}

\begin{document}

\title{Quantum Generative Models for Image Generation: Insights from MNIST and MedMNIST}

\author{
\IEEEauthorblockN{
Chi-Sheng Chen\IEEEauthorrefmark{1}\IEEEauthorrefmark{2},
Wei An Hou\IEEEauthorrefmark{1},
Hsiang-Wei Hu\IEEEauthorrefmark{1},
Zhen-Sheng Cai\IEEEauthorrefmark{1}
}
\IEEEauthorblockA{\IEEEauthorrefmark{1}Department of Artificial Intelligence in Healthcare, International Academia of Biomedical Innovation Technology, Reno, USA}
\IEEEauthorblockA{\IEEEauthorrefmark{2}Neuro Industry Research, Neuro Industry, Inc., Boston, USA \\
\{m50816m50816, weian0309, hw.hsiang.wei, s693236305\}@gmail.com}
}

\maketitle

\begin{abstract}
Quantum generative models offer a promising new direction in machine learning by leveraging quantum circuits to enhance data generation capabilities. In this study, we propose a hybrid quantum-classical image generation framework that integrates variational quantum circuits into a diffusion-based model. To improve training dynamics and generation quality, we introduce two novel noise strategies: intrinsic quantum-generated noise and a tailored noise scheduling mechanism. Our method is built upon a lightweight U-Net architecture, with the quantum layer embedded in the bottleneck module to isolate its effect. We evaluate our model on MNIST and MedMNIST datasets to examine its feasibility and performance. Notably, our results reveal that under limited data conditions (fewer than 100 training images), the quantum-enhanced model generates images with higher perceptual quality and distributional similarity than its classical counterpart using the same architecture. While the quantum model shows advantages on grayscale data such as MNIST, its performance is more nuanced on complex, color-rich datasets like PathMNIST. These findings highlight both the potential and current limitations of quantum generative models and lay the groundwork for future developments in low-resource and biomedical image generation.

\end{abstract}


\section{Introduction}
\input{sections/introduction}

\section{Related Work}
\input{sections/related_work}

\begin{figure}
    \centering
    \includegraphics[width=1\linewidth]{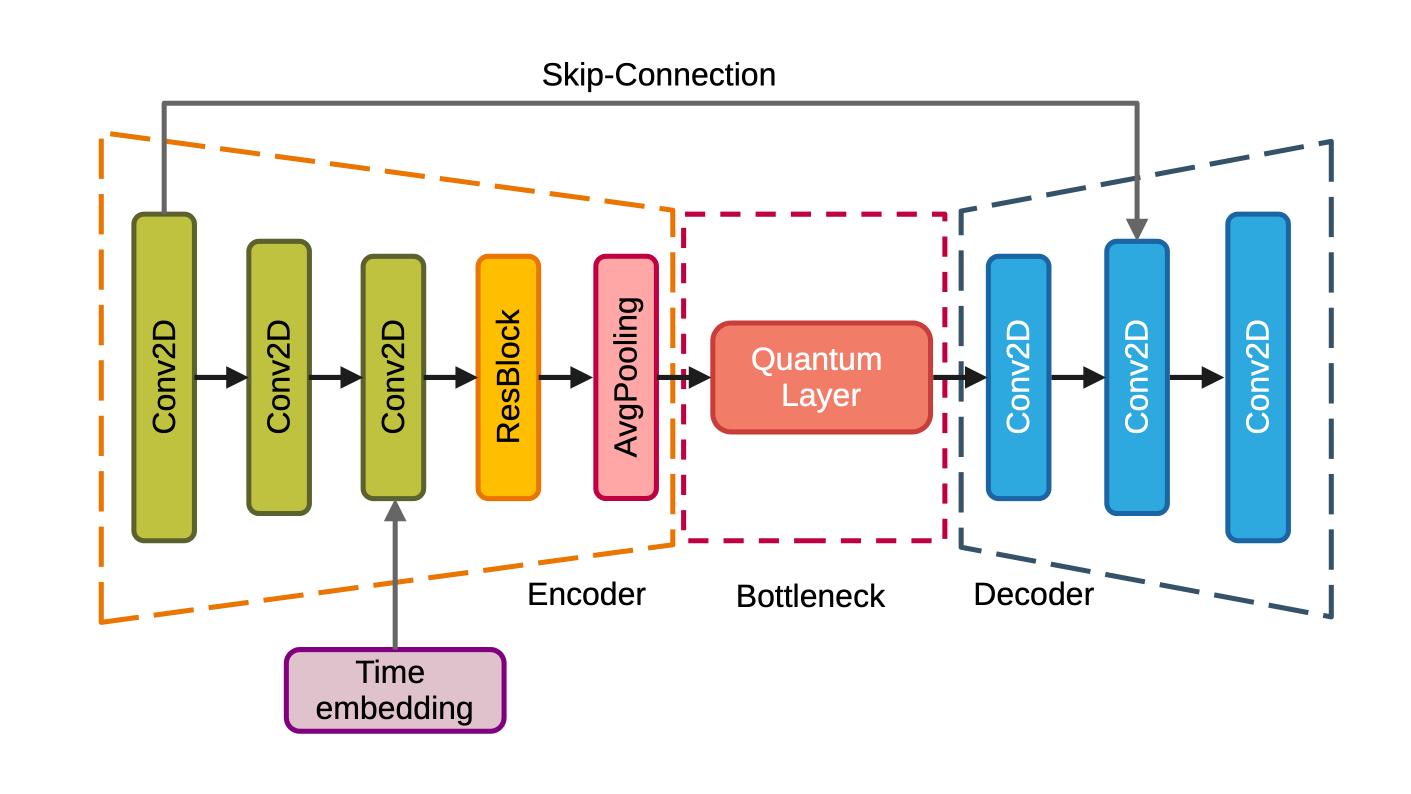}
    \caption{Model Architecture of Quantum Diffusion Model in this work.}
    \label{fig:qdiff_model}
\end{figure}

\section{Methodology}
\input{sections/methodology}

\section{Experiments and Results}
\input{sections/experiments}


\section{Discussion And Conclusion}
\input{sections/conclusion}

\bibliographystyle{IEEEtran}
\bibliography{bibliography}

\end{document}

%% file: sections/introduction.tex
Generative models have significantly advanced image synthesis in classical machine learning by enabling algorithms to learn complex data distributions and generate realistic novel samples~\cite{bond2023generalized}. Techniques such as Generative Adversarial Networks (GANs) and Variational Autoencoders (VAEs) have demonstrated impressive capabilities when trained on large-scale datasets, with applications ranging from artistic image generation to data augmentation~\cite{goodfellow2014generative}. However, as data complexity increases, classical generative models face growing challenges in terms of computational demands and model scalability.

Quantum computing introduces a novel computational paradigm for machine learning by exploiting quantum-mechanical phenomena—such as superposition and entanglement—to process information in high-dimensional Hilbert spaces. Recently, quantum generative models \cite{wall2021generative,tian2023recent} have garnered increasing attention in quantum machine learning \cite{biamonte2017quantum}, including applications in classification~\cite{chen2024qeegnet,chen2025exploring} and contrastive learning~\cite{chen2024quantum}, with the aim of achieving enhanced expressivity or computational efficiency compared to classical counterparts~\cite{lloyd2018quantum}. In particular, integrating diffusion models \cite{ho2020denoising} with quantum circuits represents a promising direction for high-fidelity image synthesis. Theoretically, parameterized quantum circuits (PQCs) can more compactly represent complex probability distributions and simulate the stochastic forward and reverse diffusion processes by leveraging the exponentially large state space of $n$ qubits.

Since this study is exploratory in nature, our primary goal is to investigate whether the inclusion of quantum layers can improve generation quality compared to classical diffusion models with identical architectures. To facilitate a fair and interpretable comparison, we design a simplified toy-model based on a lightweight U-Net architecture \cite{ronneberger2015u}, where the only modification is in the bottleneck module. Specifically, we replace the classical bottleneck with a variational quantum circuit, while keeping the encoder and decoder components unchanged. This controlled design allows us to isolate the effect of the quantum layer within the diffusion framework.

In this work, we propose a quantum diffusion model based on variational quantum circuits, implemented using the PennyLane framework. We adopt a hybrid quantum-classical architecture that integrates PQCs within the denoising process of the diffusion model, in combination with classical neural networks. Our objective is to assess the feasibility and effectiveness of quantum diffusion models on standard image synthesis benchmarks. To this end, we conduct experiments on the MNIST handwritten digit dataset \cite{deng2012mnist} and the MedMNIST biomedical image dataset \cite{yang2023medmnist}. By comparing the generated image quality and performance metrics with those of classical diffusion models, we aim to evaluate whether current quantum approaches can match or potentially surpass their classical counterparts. The remainder of this paper details related work, our proposed methodology, experimental results, and a discussion of the implications for near-term quantum hardware implementation.

%% file: sections/related_work.tex
Quantum generative models have rapidly evolved alongside advancements in quantum computing, particularly within hybrid quantum–classical frameworks. Among the foundational contributions, Quantum Generative Adversarial Networks (QGANs), introduced by Lloyd and Weedbrook, incorporate quantum circuits into adversarial training and are theoretically capable of offering exponential advantages in learning data distributions compared to their classical counterparts~\cite{lloyd2018quantum}. Subsequent experimental studies have validated the feasibility of QGAN training on noisy intermediate-scale quantum (NISQ) hardware, underscoring the practical viability of quantum adversarial learning~\cite{zoufal2019quantum}.

In practice, most implementations adopt a hybrid approach in which a quantum generator interacts with a classical discriminator, simplifying data handling while benefiting from the expressivity of quantum circuits~\cite{dallaire2018quantum}. These hybrid architectures have demonstrated promising results across various domains, such as financial distribution modeling and synthetic image generation, achieving performance comparable to classical models while requiring significantly fewer parameters in the quantum components~\cite{zoufal2019quantum, situ2020quantum}.

Quantum Variational Autoencoders (QVAEs) incorporate quantum sampling mechanisms into classical VAE frameworks. Early work has shown that using quantum Boltzmann machines to model latent spaces yields superior performance in reconstructing and generating MNIST digits~\cite{khoshaman2018quantum}. Likewise, Quantum Circuit Born Machines (QCBMs) utilize parameterized quantum circuits to represent probability distributions and have been effective in learning binary patterns and small-scale images using optimization techniques such as Maximum Mean Discrepancy~\cite{liu2018differentiable}.

In the domain of medical imaging, recent research has begun to explore quantum generative models specifically tailored for biomedical datasets. For example, Quantum Image Generative Learning (QIGL) models integrate quantum circuits with dimensionality reduction techniques like principal component analysis (PCA) to efficiently generate medically relevant images. These models have shown promise in synthesizing realistic medical patches, offering potential utility for data augmentation in disease classification tasks~\cite{khatun2023quantum}. Another notable line of work has examined quantum-enhanced classification on the MedMNIST benchmark, demonstrating both the potential and limitations of purely quantum classifiers when applied to realistic medical imaging data on current quantum processors~\cite{singh2023benchmarking}.

Recently, diffusion models have surpassed GANs in generating higher-quality images across many classical generative tasks \cite{croitoru2023diffusion}, inspiring a shift in focus toward diffusion-based frameworks. Motivated by these advances, and recognizing the current lack of research on quantum diffusion models image generation related tasks \cite{kolle2024quantum, parigi2024quantum}, this work aims to conduct an exploratory investigation into their potential. Specifically, we examine whether incorporating quantum circuits into diffusion models—guided by insights from their classical counterparts—can offer practical benefits in image synthesis.


%% file: sections/methodology.tex
\subsection{Overview}

We propose a hybrid quantum-classical diffusion model based on a simplified U-Net architecture with an optional variational quantum attention layer. Our goal is to investigate the impact of incorporating a parameterized quantum circuit (PQC) within the bottleneck of a diffusion-based image generation pipeline. Both classical and quantum-enhanced models are implemented using the same structure and training procedures to ensure a fair comparison. The experiments are performed on class-conditional subsets of the MNIST dataset.

\subsection{Diffusion Process}

We adopt the standard forward diffusion process $q(x_t \mid x_0)$, where Gaussian noise is progressively added to the original image. The process is defined as:
\begin{equation}
q(x_t \mid x_0) = \mathcal{N}(x_t; \sqrt{\bar{\alpha}_t}x_0, (1 - \bar{\alpha}_t)\mathbf{I}),
\label{eq:forward_diffusion}
\end{equation}

where $\bar{\alpha}_t = \prod_{s=1}^{t}(1 - \beta_s)$. We use a cosine schedule~\cite{nichol2021improved} to compute the $\beta_t$ values for smoother training dynamics:
\begin{equation}
\bar{\alpha}_t = \cos^2\left(\frac{t/T + s}{1 + s} \cdot \frac{\pi}{2}\right),
\label{eq:cosine_schedule}
\end{equation}
\begin{equation}
\beta_t = 1 - \frac{\bar{\alpha}_t}{\bar{\alpha}_{t-1}}.
\label{eq:beta_schedule}
\end{equation}

The model $f_\theta$ is trained to predict the noise $\epsilon$ added at each time step $t$. The loss function is defined as:
\begin{equation}
\mathcal{L}_{\text{MSE}} = \mathbb{E}_{x_0, \epsilon, t}\left[ \left\| f_\theta(x_t, t) - \epsilon \right\|_2^2 \right].
\label{eq:mse_loss}
\end{equation}

\subsection{Improved U-Net Architecture and Timestep Embedding}

Our denoising model is a lightweight U-Net composed of three convolutional encoder layers, a residual block at the bottleneck, and two transposed convolutional decoder layers. Given an input $x \in \mathbb{R}^{1 \times 28 \times 28}$ and timestep $t$, we embed the timestep using sinusoidal positional encoding:
\begin{equation}
\gamma(t) = \left[\sin(t \cdot \omega_k), \cos(t \cdot \omega_k)\right]_{k=1}^{d/2}, \quad \omega_k = 10000^{-2k/d},
\label{eq:sinusoidal_embedding}
\end{equation}
\begin{equation}
\text{Embed}(t) = \text{Linear}(\gamma(t)) \in \mathbb{R}^{128}.
\label{eq:time_embedding}
\end{equation}

This embedding is projected and added to the intermediate feature maps to condition the model on diffusion step $t$.

\begin{figure}
    \centering
    \includegraphics[width=1\linewidth]{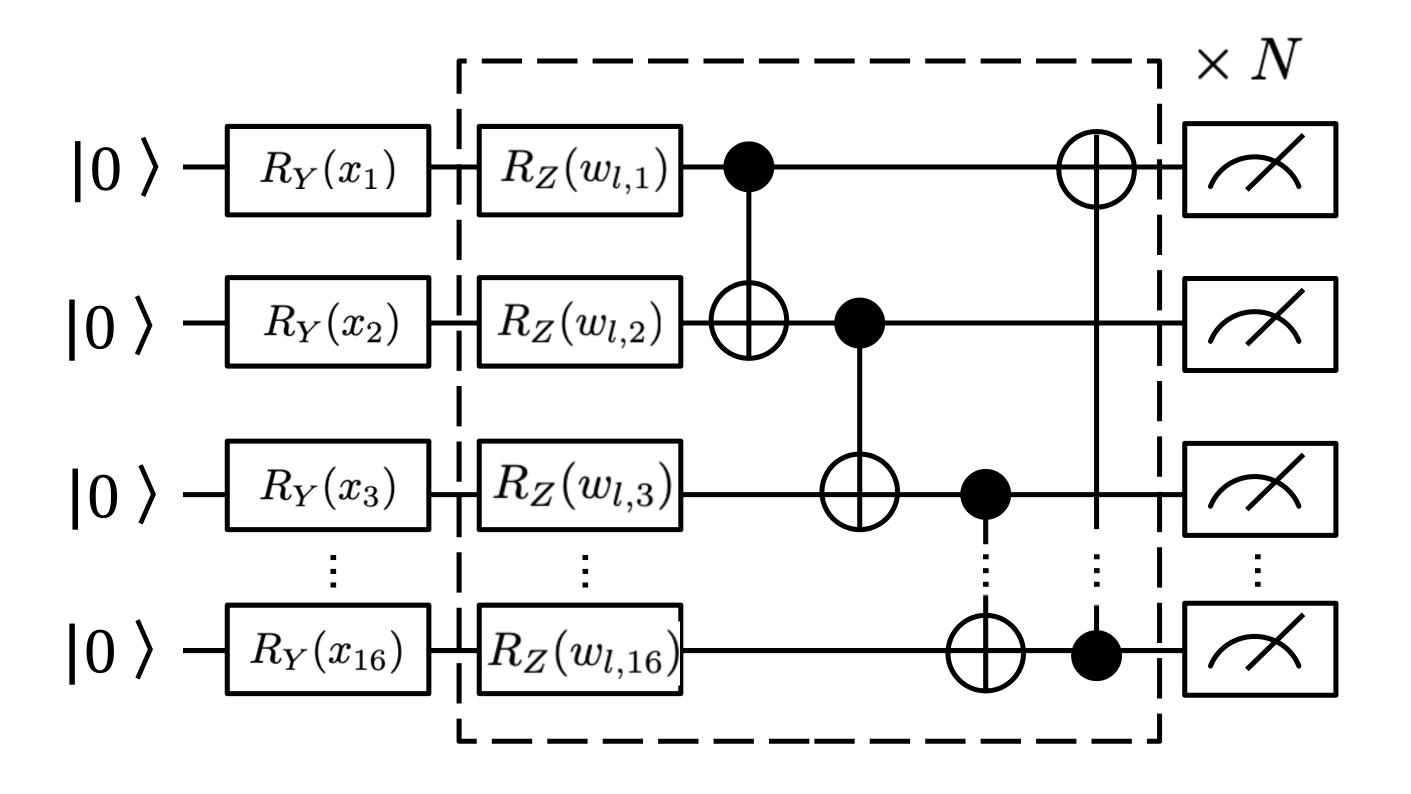}
    \caption{The VQC (quantum layer) used in this work.}
    \label{fig:qnn}
\end{figure}

\subsection{Quantum Layer}

In the quantum-enhanced version of our model, we integrate a variational quantum attention layer into the bottleneck of the U-Net architecture like Figure~\ref{fig:qnn}. Specifically, the bottleneck feature map is first processed via global average pooling to obtain a latent vector \( z \in \mathbb{R}^{128} \). This vector is then linearly projected into a 16-dimensional space, which serves as the input to the quantum circuit.

Each component of the resulting 16-dimensional vector is encoded into a quantum state using single-qubit \( \text{RY}(x_i) \) rotation gates applied to \( n = 16 \) qubits. The quantum circuit consists of \( L = 3 \) variational layers. Within each layer, trainable \( \text{RZ}(w_{l,i}) \) gates are applied to each qubit, followed by a series of \( n-1 \) CNOT gates that create a linear entanglement topology across adjacent qubits. After passing through the variational layers, the quantum circuit outputs a vector of Pauli-Z expectation values: $R_Z(w_{l,16})$

\begin{equation}
z_{\text{quantum}} = \left[ \langle Z_i \rangle \right]_{i=1}^{16} \in \mathbb{R}^{16}.
\label{eq:quantum_expectation}
\end{equation}

This quantum output is projected back to \( \mathbb{R}^{128} \) using a fully connected layer and reshaped to dimensions \( 128 \times 1 \times 1 \), allowing it to be broadcast-multiplied elementwise with the original bottleneck feature map. The resulting quantum-enhanced representation is thus computed as:

\begin{equation}
x_{\text{quantum}} = x_{\text{bottleneck}} \odot \tilde{z}.
\label{eq:quantum_attention}
\end{equation}

The entire quantum module is implemented using PennyLane as a differentiable QNode. It is fully integrated into the PyTorch training pipeline and supports gradient-based optimization through the parameter-shift rule~\cite{schuld2019evaluating}, enabling end-to-end training alongside classical model components.

\subsection{Training Strategy and EMA}

The model is trained on individual classes from MNIST and MedMNIST. At each training step, a random timestep $t \sim \mathcal{U}(0, T)$ is sampled, and the input image $x_0$ is perturbed with Gaussian noise:
\begin{equation}
x_t = \sqrt{\bar{\alpha}_t}x_0 + \sqrt{1 - \bar{\alpha}_t} \cdot \epsilon,
\quad \epsilon \sim \mathcal{N}(0, \mathbf{I}).
\label{eq:forward_sampling}
\end{equation}

The model predicts the added noise $\hat{\epsilon}$, and the mean squared error loss is computed as in Eq.~\eqref{eq:mse_loss}. Optimization is performed using Adam with a learning rate of $3 \times 10^{-4}$. To stabilize training and improve sample quality, we apply exponential moving average (EMA) to the model parameters:
\begin{equation}
\theta_{\text{EMA}} \leftarrow \beta \cdot \theta_{\text{EMA}} + (1 - \beta) \cdot \theta, \quad \beta = 0.999.
\label{eq:ema}
\end{equation}

After each epoch, samples are generated using the EMA model and saved for visual inspection. The best-performing model is retained for inference.

%% file: sections/experiments.tex
\subsection{Datasets and Experimental Setup}
We evaluate our proposed quantum generative model on two datasets: MNIST and MedMNIST. MNIST is a well-known dataset consisting of $28\times28$ grayscale images of handwritten digits (0 to 9), with 60,000 training samples and 10,000 test samples. To simplify the quantum learning task, we use a subset of MNIST (e.g., digits 0 and 1) for training. Nonetheless, our model architecture is, in principle, capable of handling all digits provided sufficient model capacity. MedMNIST is a collection of biomedical image datasets with a structure similar to MNIST~\cite{yang2023medmnist}, where each sub-dataset contains $28\times28$ images tailored for specific medical imaging tasks (e.g., ChestMNIST for chest X-ray lung nodule classification, PathMNIST for pathology images, etc.), with associated class labels. We select one of the 2D MedMNIST subsets (e.g., PathMNIST) as a test case for complex image generation to evaluate the model's generalization in more challenging domains.

In our implementation, both training and generation are performed using the original image resolution, without any additional downsampling, dimensionality reduction, or padding. The MNIST grayscale images are normalized to the range $[-1, 1]$ and treated as single-channel inputs. In contrast, the PathMNIST images retain their original RGB format, and each channel is independently normalized. The discriminator is implemented as a standard convolutional neural network (CNN) that accepts full $28\times28$ images without requiring interpolation or resizing. The quantum module operates only on the intermediate latent feature vector (the bottleneck representation) and adjusts attention through a parameterized quantum circuit; therefore, it does not impose additional constraints on the input or output image dimensions.

For training, we use a unified setting of 30 epochs for both MNIST and PathMNIST datasets to ensure fair comparisons between the classical and quantum-enhanced models. The batch size is fixed at 64, and we use the Adam optimizer with a learning rate of $3\times10^{-4}$. Unlike traditional QGAN architectures that require repeated quantum circuit executions (shots) to estimate expectation values, our quantum attention module is embedded directly into the PyTorch computation graph. Due to the current limitations of quantum hardware, we simulate the quantum computation using a quantum simulator backend. This setup supports end-to-end automatic differentiation, allowing the model to output continuous vectors in a single forward pass without the need for repeated sampling, which facilitates stable and efficient training.

\subsection{Evaluation Metrics}

To quantitatively assess the quality of images produced by the quantum generative model, we computed several standard evaluation metrics used in generative modeling. The Fréchet Inception Distance (FID) \cite{heusel2017gans} measures the distance between two multivariate Gaussians fitted to features extracted from a pretrained Inception network on real and generated images, respectively. Formally, let $\mathcal{N}(\mu_r, \Sigma_r)$ and $\mathcal{N}(\mu_g, \Sigma_g)$ be the feature distributions of real and generated images, then the FID is defined as:

\begin{equation}
\text{FID} = \|\mu_r - \mu_g\|^2 + \text{Tr}(\Sigma_r + \Sigma_g - 2(\Sigma_r \Sigma_g)^{1/2}).
\label{eq:fid}
\end{equation}

Lower FID values indicate closer similarity between the distributions, and hence better generative quality.

We also compute the Structural Similarity Index (SSIM) \cite{wang2004image} between generated images $x_g$ and reference real images $x_r$ to assess perceptual quality. SSIM is defined as:

\begin{equation}
\text{SSIM}(x_r, x_g) = \frac{(2\mu_r \mu_g + C_1)(2\sigma_{rg} + C_2)}{(\mu_r^2 + \mu_g^2 + C_1)(\sigma_r^2 + \sigma_g^2 + C_2)},
\label{eq:ssim}
\end{equation}

where $\mu_r$, $\mu_g$ are the mean pixel intensities, $\sigma_r^2$, $\sigma_g^2$ the variances, $\sigma_{rg}$ the covariance, and $C_1$, $C_2$ are constants to stabilize the division.

For MNIST and MedMNIST, although the datasets are primarily designed for classification tasks, we do not rely on classification accuracy for evaluation. Instead, since our model is a diffusion-based generator, we directly optimize and monitor the denoising performance during training using the mean squared error (MSE) between the predicted and true noise components.

The training objective is based on the standard denoising diffusion probabilistic model (DDPM) framework, where the generator $f_\theta$ is trained to estimate the additive noise $\epsilon$ at each timestep $t$ for a noisy input $x_t$. The loss function is defined as:

\begin{equation}
\mathcal{L}_G = \mathbb{E}_{x_0, t, \epsilon}\left[\left\| f_\theta(x_t, t) - \epsilon \right\|^2 \right],
\label{eq:generator_loss_diffusion}
\end{equation}

where $x_t = \sqrt{\bar{\alpha}_t} x_0 + \sqrt{1 - \bar{\alpha}_t} \epsilon$ represents the noisy sample at diffusion step $t$, and $\bar{\alpha}_t$ is the cumulative product of $\alpha_t = 1 - \beta_t$ over $t$ steps. The model is trained to reverse this forward process.

During training, we also apply exponential moving average (EMA) on the generator weights to stabilize the training dynamics and improve sampling quality. The EMA model is used to generate samples for visualization and evaluation after each epoch.

We do not employ an explicit discriminator or adversarial loss; instead, the quality of generation is evaluated by tracking the training loss, visual inspection of sampled images, and computing perceptual similarity metrics such as SSIM and FID against real images.

\begin{table*}[htbp]
\centering
\caption{Comparison of SSIM and FID between Classical and Quantum Diffusion Models on MNIST Digits}
\begin{tabular}{l cccccccccc c c}
\hline
\textbf{Model} & \textbf{0} & \textbf{1} & \textbf{2} & \textbf{3} & \textbf{4} & \textbf{5} & \textbf{6} & \textbf{7} & \textbf{8} & \textbf{9} & \textbf{Avg} & \textbf{Std} \\
\hline
\multicolumn{13}{c}{\textbf{SSIM (Higher is better)}} \\
\hline
Classical & 0.0945 & 0.1494 & 0.1165 & 0.1093 & 0.1119 & 0.0685 & 0.1238 & 0.1082 & 0.1246 & 0.0787 & 0.1085 & 0.0234 \\
Quantum   & 0.1241 & 0.1153 & 0.0961 & 0.0752 & 0.2000 & 0.0603 & 0.2052 & 0.1126 & 0.1228 & 0.1513 & \textbf{0.1263} & 0.0477 \\
\hline
\multicolumn{13}{c}{\textbf{FID (Lower is better)}} \\
\hline
Classical & 218.61 & 261.67 & 310.63 & 219.85 & 259.43 & 313.99 & 325.59 & 293.78 & 255.38 & 251.58 & 271.05 & 38.18 \\
Quantum   & 268.96 & 226.65 & 236.24 & 325.95 & 259.93 & 269.04 & 270.39 & 280.68 & 205.28 & 249.38 & \textbf{259.25} & 33.04 \\
\hline
\label{tab:mnist_results}
\end{tabular}
\end{table*}

\begin{table}[htbp]
\centering
\caption{SSIM and FID Comparison on PathMNIST}
\begin{tabular}{l c c}
\hline
\textbf{Model} & \textbf{SSIM} & \textbf{FID} \\
\hline
Classical & \textbf{0.4107} & 95.72 \\
Quantum   & 0.0931 & \textbf{84.40} \\
\hline
\label{tab:pathmnist_results}
\end{tabular}
\end{table}

\begin{figure}[htbp]
    \centering
    \begin{subfigure}[b]{1\linewidth}
        \centering
        \includegraphics[width=\linewidth]{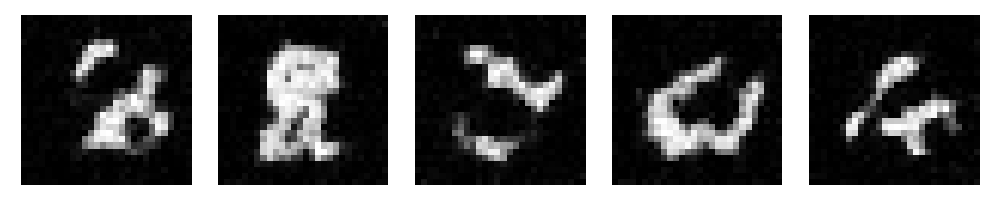}
        \caption{Generated MNIST digit 0 image from the classical diffusion model.}
    \end{subfigure}
    \vspace{0.5em}
    \begin{subfigure}[b]{1\linewidth}
        \centering
        \includegraphics[width=\linewidth]{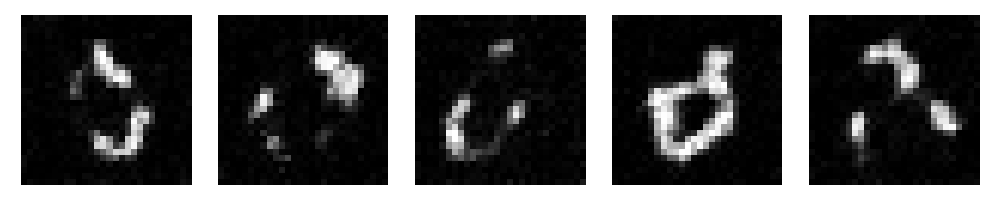}
        \caption{Generated MNIST digit 0 image from the quantum diffusion model.}
    \end{subfigure}
    \caption{Comparison between classical and quantum diffusion models on MNIST digit 0.}
    \label{fig:mnist_digit0_comparison}
\end{figure}

\subsection{Results on MNIST}
\begin{figure}[htbp]
    \centering
    \begin{subfigure}[b]{1\linewidth}
        \centering
        \includegraphics[width=\linewidth]{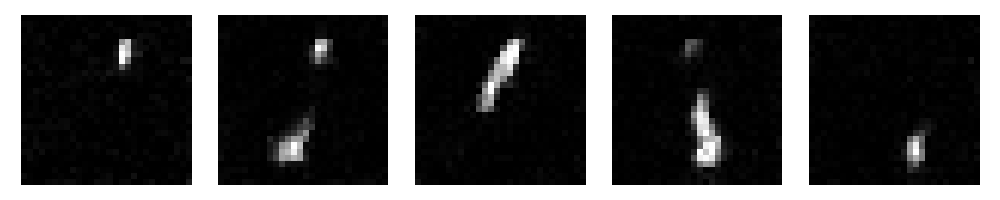}
        \caption{Generated MNIST digit 1 image from the classical diffusion model.}
    \end{subfigure}
    \vspace{0.5em}
    \begin{subfigure}[b]{1\linewidth}
        \centering
        \includegraphics[width=\linewidth]{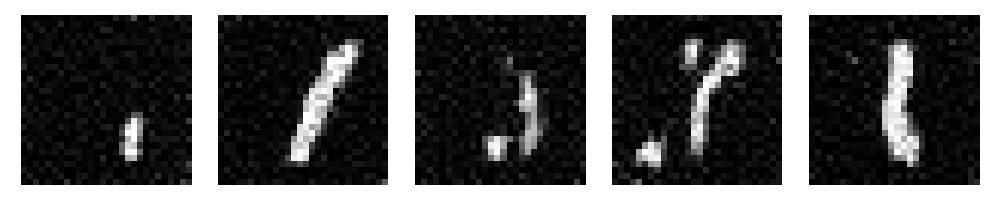}
        \caption{Generated MNIST digit 1 image from the quantum diffusion model.}
    \end{subfigure}
    \caption{Comparison between classical and quantum diffusion models on MNIST digit 1.}
    \label{fig:mnist_digit1_comparison}
\end{figure}

\begin{figure}[htbp]
    \centering
    \begin{subfigure}[b]{1\linewidth}
        \centering
        \includegraphics[width=\linewidth]{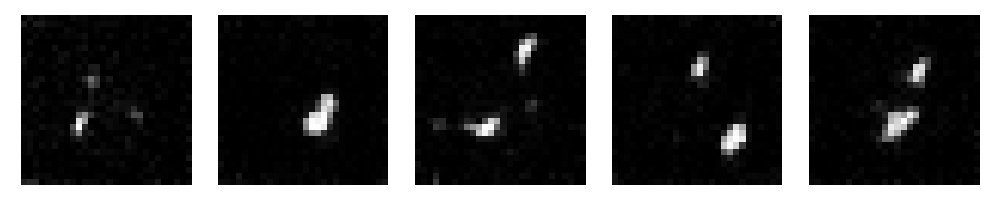}
        \caption{Generated MNIST digit 6 image from the classical diffusion model.}
    \end{subfigure}
    \vspace{0.5em}
    \begin{subfigure}[b]{1\linewidth}
        \centering
        \includegraphics[width=\linewidth]{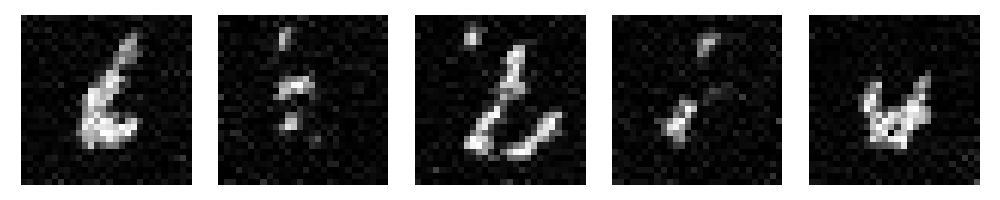}
        \caption{Generated MNIST digit 6 image from the quantum diffusion model.}
    \end{subfigure}
    \caption{Comparison between classical and quantum diffusion models on MNIST digit 6.}
    \label{fig:mnist_digit6_comparison}
\end{figure}

\begin{figure}[htbp]
    \centering
    \begin{subfigure}[b]{1\linewidth}
        \centering
        \includegraphics[width=\linewidth]{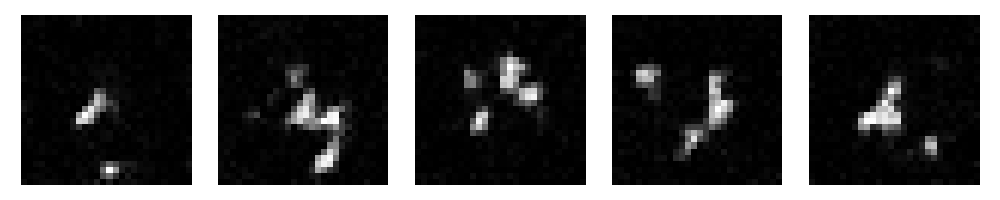}
        \caption{Generated MNIST digit 9 image from the classical diffusion model.}
    \end{subfigure}
    \vspace{0.5em}
    \begin{subfigure}[b]{1\linewidth}
        \centering
        \includegraphics[width=\linewidth]{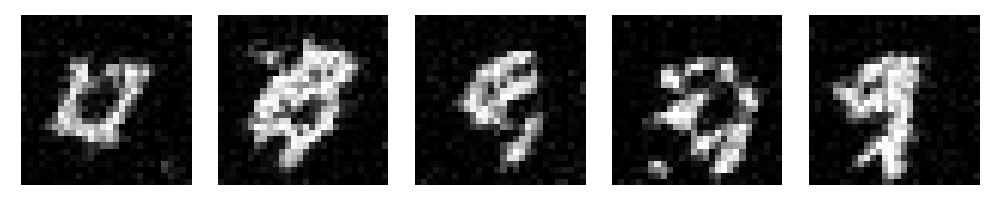}
        \caption{Generated MNIST digit 9 image from the quantum diffusion model.}
    \end{subfigure}
    \caption{Comparison between classical and quantum diffusion models on MNIST digit 9.}
    \label{fig:mnist_digit9_comparison}
\end{figure}

To evaluate the performance of our quantum diffusion model, we conducted experiments on the MNIST dataset and compared the results with a classical diffusion baseline. Figure~\ref{fig:mnist_digit0_comparison}, Figure~\ref{fig:mnist_digit1_comparison}, Figure~\ref{fig:mnist_digit6_comparison} and Figure~\ref{fig:mnist_digit9_comparison} show a comparison of the images generated by the diffusion model after training for 30 epochs. Two widely adopted image quality metrics were used for evaluation: Structural Similarity Index (SSIM) and Fréchet Inception Distance (FID). The SSIM metric reflects perceptual image quality (higher is better), while FID assesses the distributional similarity between generated and real images (lower is better).

Table~\ref{tab:mnist_results} summarizes the SSIM and FID scores across all digit classes (0–9), as well as their average and standard deviation.

For SSIM, the quantum model achieved a higher average score of 0.1263 compared to the classical model's 0.1085, indicating superior perceptual quality in the generated digits. Notably, the quantum model outperformed the classical model on several digits, such as digit 4 (0.2000 vs. 0.1119) and digit 6 (0.2052 vs. 0.1238), demonstrating the effectiveness of quantum layers in enhancing structural features. However, it exhibited a higher standard deviation (0.0477), suggesting slightly more variability in performance across different digits.

In terms of FID, the quantum model also showed a modest improvement with a lower average score of 259.25 compared to 271.05 for the classical model, implying better alignment between the generated and real image distributions. The quantum model achieved noticeably lower FID scores on certain digits, such as digit 1 (226.65 vs. 261.67) and digit 8 (205.28 vs. 255.38). The standard deviation of the FID scores was also lower for the quantum model (33.04 vs. 38.18), indicating more consistent performance.

These results suggest that incorporating quantum components into the diffusion process can enhance both the fidelity and perceptual quality of generated images, particularly in capturing fine-grained structural details.

\subsection{Results on MedMNIST}
\begin{figure}[htbp]
    \centering
    \begin{subfigure}[b]{1\linewidth}
        \centering
        \includegraphics[width=\linewidth]{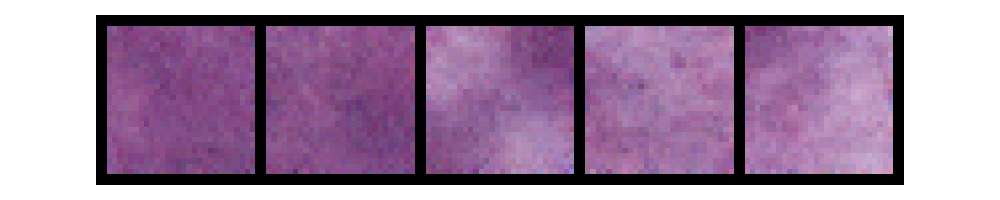}
        \caption{Generated PathMNIST image from the classical diffusion model.}
    \end{subfigure}
    \vspace{0.5em}
    \begin{subfigure}[b]{1\linewidth}
        \centering
        \includegraphics[width=\linewidth]{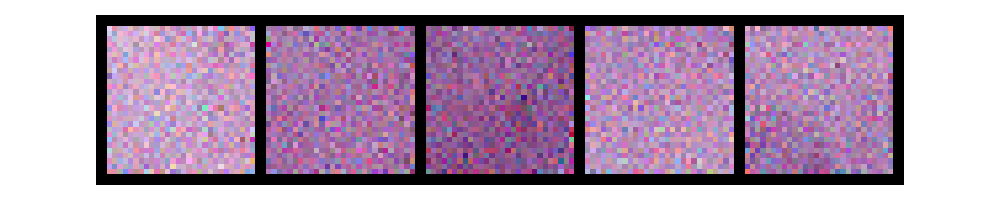}
        \caption{Generated PathMNIST image from the quantum diffusion model.}
    \end{subfigure}
    \caption{Comparison between classical and quantum diffusion models on PathMNIST.}
    \label{fig:mnist_pathmnist_comparison}
\end{figure}

We further evaluated the performance of our diffusion models on the PathMNIST dataset, which contains more complex and color-rich medical images compared to MNIST. Figure~\ref{fig:mnist_pathmnist_comparison} show a comparison of the images generated by the diffusion model after training for 30 epochs. The results, summarized in Table~\ref{tab:pathmnist_results}, show a more nuanced trade-off between SSIM and FID.

The classical diffusion model achieved a significantly higher SSIM score of 0.4107 compared to the quantum model's 0.0931, suggesting that the classical model better preserved the structural similarity in the generated color images. This indicates that the quantum model struggled to capture fine-grained spatial details and color consistency in high-dimensional visual content.

However, the quantum model achieved a better FID score of 84.40 versus 95.72 for the classical model, demonstrating superior performance in terms of matching the global data distribution of real images. This may imply that while the quantum model generates samples that statistically align more closely with the real dataset, the individual image quality and perceptual similarity are less satisfactory in the context of complex, high-resolution color images.

These results suggest that the benefits of quantum diffusion models observed in grayscale datasets like MNIST may not directly transfer to more complex, color-based datasets like PathMNIST. Future work may focus on improving quantum architectures to better handle color information and spatial complexity.

%% file: sections/conclusion.tex
In this work, we proposed a variational quantum circuit-based generative diffusion model for image synthesis, implemented using the PennyLane hybrid quantum-classical framework. Through empirical evaluation on both MNIST and PathMNIST datasets, we explored the strengths and limitations of quantum diffusion-based generative approaches compared to classical baselines.

Our experiments reveal that the quantum model performs competitively on simpler grayscale datasets like MNIST, achieving higher average SSIM and lower FID in some cases, particularly in capturing fine-grained digit structures. Notably, under low-data regimes—when trained with fewer than 100 samples—the quantum model consistently outperformed the classical counterpart using the same architecture, indicating superior generalization ability and robustness in data-scarce settings. These findings suggest that quantum circuits can effectively encode structural patterns within compact representations, even under current hardware constraints.

However, this performance advantage does not fully carry over to more complex and high-resolution color datasets. On PathMNIST, the quantum model achieved a better FID score, indicating improved alignment with the global data distribution, but suffered from significantly lower SSIM values, reflecting a lack of fine-grained spatial fidelity in generated images.

This discrepancy highlights a key limitation: while quantum models are capable of capturing global distributional properties, they may struggle with preserving detailed local features—especially in high-dimensional, color-rich image domains. This suggests that current quantum circuit designs may be better suited for learning abstract or low-dimensional representations rather than detailed spatial structures, which are crucial in domains such as medical imaging.

Despite these limitations, our study presents one of the first applications of quantum generative modeling on biomedical image datasets like MedMNIST. We demonstrate the feasibility of training and evaluating quantum generators with classical feedback loops, and provide a practical framework for comparing quantum and classical diffusion models under standardized metrics.

Looking ahead, future research should explore deeper and more expressive quantum circuit ansätze, increase qubit counts, and investigate hybrid strategies that combine quantum feature extraction with classical post-processing. Alternative generative paradigms, such as quantum-enhanced autoencoders and quantum diffusion models with learnable noise schedules, may offer improvements in both stability and generation quality. Finally, deploying these models on real quantum hardware will be essential to assess their true potential and practical limitations.

As quantum technology continues to advance, we believe that hybrid quantum-classical generative models will become increasingly relevant in fields requiring efficient, compact, and interpretable data generation—ranging from secure data synthesis to low-resource medical imaging applications.